**Strategies for fostering Knowledge Management Programs in Public Organizations**

Hugo A. Mitre-Hernández[a], Arturo Mora-Soto[a], Héctor Pérez López-Portillo[b] and Carlos Lara-Alvarez[a]
[a]Center for Research in Mathematics (CIMAT, A.C.). Zacatecas, Mexico
[b]University of Guanajuato. Guanajuato, Mexico.
hmitre@cimat.mx
arturo.mora@cimat.mx
h.perezlopezportillo@ugto.mx
carlos.lara@cimat.mx

*Abstract*: *Knowledge Management* (KM) is an approach to achieving strategic objectives by visualizing, sharing, and using intangible resources of an organization and its stakeholders. There are many studies that analyze specific factors for the successful implementation of KM programs, and the evaluation of such factors is considered a strategic tool for Public Organizations (POs) for efficiently directing the implementation of a KM program. Nevertheless, there are cultural problems such as weak trust, bad collaboration; technological problems as KM systems difficult to use, nor interconnected or interoperable; and strategic problems as political changes, not inter-administration continuity or lack of political willingness. In this research we provide an overview of the key factors for facilitating the implementation of KM programs. To this end, we conducted a Systematic Literature Review (SLR) of success factors to create a KM program for POs. Analyzing the twenty related studies we summarize benefits and quality attributes of KM. Moreover, we obtained from an ongoing project evaluation results of the KM factors ranked cultural, technological and strategic. The results show that the Mexican POs have strategic and technological issues such as: a misalignment between knowledge management and organizational goals, it seems that POs are barely reusing their knowledge to execute their daily activities or to take decisions, and is not possible to know who and how a knowledge asset has been used. Finally, according to the gaps and difficulties of the SLR, we provide strategies to successfully implement KM programs in Mexico.

## 1. INTRODUCTION

The main objective of Knowledge Management (KM) in public sector is to improve the effectiveness and viability of the knowledge domain of a Public Organization (POs) (Mbhalati 2014). A general goal of KM is *"to improve the systematic handling of knowledge and potential knowledge within the organization* (Heisig 2009, p.5)"*, and according to Tsui et al. (2009, p.986), the purpose of KM is to provide online, real time access to knowledge throughout the organization and to its customers which act as enablers and catalysts for innovative application in government agencies (Tsui et al. 2009, p.988)". Therefore KM is seen as an effective solution that can support public administrative activities of modernizing government.

From the institutional point of view, the benefits of KM implementation are the cost reduction, whilst promoting economic development, increasing transparency in government, improving service delivery and facilitating the advancement of an information society (The World Bank 2011; United Nations 2014). In the research area, authors recognize the importance of assume preview factors for leading the initiatives of KM in POs (Syed-Ikhsan & Rowland 2004), and mostly of KM enablers included managerial influences, job expertise, social relationships (Pee & Kankanhalli 2008) and maximize productivity in the public sector across different levels of government (Jain 2009).

Government organizations are major knowledge consumers and producers (Jain 2009), KM has been encouraged by POs intention to upgrade and modernize its core activities (Braun & Mueller 2014). KM programs in POs focus on ways to manage and distribute what government institutions know internally and among other POs, with the purpose of taking collaborative decisions (Jennex & Smolnik 2011). This management breakthrough can contribute to a more efficient, transparent and social needs sensible acting from government (De Angelis 2013). Consequently, KM is crucial for institutional success but for societal development (Ragab & Arisha 2013).

From a cultural approach KM in POs can be seen as a strategy for recover trust in government (United Nations 2007), its adoption involves innovation and sometimes implies reforming institutions. As we said before, KM can contribute to fight against corruption practices (Elkadi 2013; Tung & Rieck 2005), for that reason a major public aware of KM benefits is needed (Tung & Rieck 2005, p.437).

Even with these benefits mentioned above, there is a lack of awareness of KM in the public sector. For an effective inclusion of KM model to assist governments develop and capitalize knowledge-based stakeholder partnerships through knowledge management programs, is essential to assess cultural, technological and strategic factors. We found, in the academic literature review some difficulties and shortcomings of KM in POs classified in cultural (Massingham & Massingham 2014), technological (Jain 2009) and strategic (Liberona & Ruiz 2013) domain. With the purpose of avoiding those problems, it is critical to perform a preliminary evaluation of KM factors. This is the main reason why this research has been focused on KM factors.

The objectives of this work are:
(I) To summarize benefits and quality attributes (understandability, clarity and credibility) of KM in POs through a Systematic Literature Review (SLR).
(II) To obtain the benefits, difficulties, evaluations and proposals from KM studies in a systematic way and to identify cultural, strategic and technological strategies to foster KM Programs in Mexican POs.

In order to accomplish the first objective, authors conducted a systematic literature review (SLR) to explore the state of the art in knowledge management in the public sector. For the second objective, we obtained from an ongoing project evaluation results of the KM factors ranked cultural, technological and strategic applied to five Mexican POs. Moreover, by considering the results of the SLR in terms of benefits, contributions and the current diagnosis of the maturity of KM management in Mexican POs, it is possible to recommend strategies to implement KM programs in the Mexican public sector.

The rest of the article is organized as follows: Section 2. Describes the SLR; It summarizes the benefits and quality factors of understandability, clarity and credibility of KM in POs. Section 3. Discusses the strategies to enable KM programs in POs. Finally, Section 4 present our discussions.

## 2. SYSTEMATIC LITERATURE REVIEW
The goal of this Systematic Literature Review (SLR) is to explore the state of the art in managing knowledge in the public sector. With the guidelines of the SLR by (Kitchenham & Charters 2007; Petersen et al. 2008), we intended at searching, selecting, evaluating, and presenting studies to respond to research questions (see figure 2).

The resulting studies of the research questions organized in cultural, technological and strategic dimensions, allowed us to discover a set of activities, which emphasize as a specific domain that is an object of interest, we call it knowledge domain of KM factors in public organizations. As been seen in figure 1, such activities of the knowledge domain are modeled in a process. The modeled process is similar to the KM processes formulated by Lee et al. (2005) and Ding et al. (2014). The following paragraphs describe each activity:

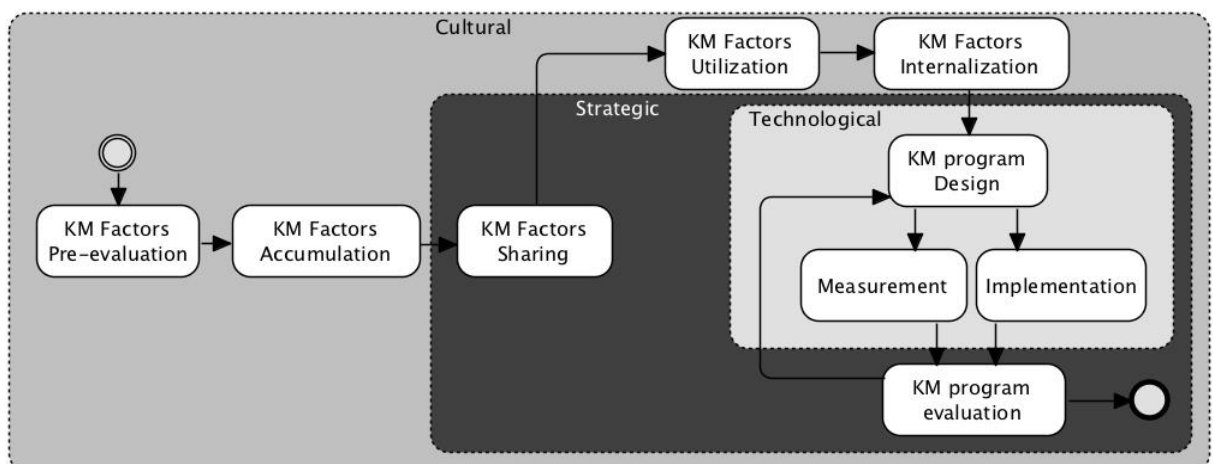
Figure 1: SLR scope in a process modeled

The following four activities are immerse on ***cultural dimension***, characterized by trust, collaboration and openness (Alsadhan et al. 2008), a culture that supports KM is one that values knowledge and encourages its creation, sharing and application (Wong 2005). *A collaboration environment provides opportunities for knowledgeable people to share knowledge openly and have successful knowledge management programs* (Allameh et al. 2011, p.1217).

- **KM factors Pre-evaluation:** The aim of this activity is to better understand the knowledge resources that are available, in order to identify what knowledge is needed, why it is needed, and who is likely to have this knowledge. Seen as K-map, helps an organization to visualize the relationships and processes that connect their knowledge resources, which include people, documentation (Chong & Chong 2009, p.147), and K-audit that helps organizations to identify where knowledge resides within their firms (Chong & Chong 2009, p.147).
- **KM factors accumulation:** This activity is related to understanding of personally acquired information and employees' search through corporate databases to obtain knowledge required for given tasks (Ho et al. 2014). Saving the knowledge of organization is one of the most important elements of a KM system (Akhavan et al. 2006, p.109).
- **KM factors utilization:** this activity facilitates tasks by applying knowledge, is about the degree of knowledge used by the organization and the knowledge (Ho et al. 2014; Lee et al. 2005).
- **KM factor internalization:** this activity produces new knowledge from learning and applying best practices (Ho et al. 2014).

The next two factors are commonly associated with the ***strategic dimension*** due the fundamental alignment that may occur between KM strategies with organizational core objectives (Jennex & Smolnik 2011, chap.5).

- **KM factors sharing:** this activity promotes diffusion of knowledge and contributes to making the work process astute and knowledge-intensive (Lee et al. 2005; Ho et al. 2014). Sharing plays an important role on implementing and executing KM system, knowledge sharing between employees needs a strong culture, trust and also transparency in all over the organization (Akhavan et al. 2006), leadership, time allocation, and trust encourage knowledge sharing (Mas-Machuca & Martínez Costa 2012), and social interaction, rewards, and organizational support had a significant effect on sharing knowledge (Amayah 2013). Without sharing, it is almost impossible for knowledge to be transferred to another person (Syed-Ikhsan & Rowland 2004).
- **KM program evaluation:** It has been demonstrated how measurement is an important step in KM (Wong et al. 2013). This activity seeks to entirely assess KM program implementation in order to observe what is working or not within KM program, and therefore make an well-versed judgment to adjust (Andone 2009). Also helps to support managers' decisions on KM strategy to improve KM process performance within the organization (Bose & Ranjit 2004; Wong et al. 2013).

These three activities are in ***technological dimension*** because they influence each other working as infrastructural drivers that enable KM programs in POs (Syed-Ikhsan & Rowland 2004), and also because there are activities commonly associated within this dimension (Mas-Machuca & Martínez Costa 2012).

- **KM program design:** this activity creates a KM project and establishes its goals and objectives to facilitate organizational processes. It can be seen as propose guidelines for solving specific problems of linking knowledge and organizational processes, better usage and provision of knowledge when and where it is needed. The aim of this activity is the complete conceptualization of KM programs; design encompasses all KM barriers and enablers (Allameh et al. 2011; Pee & Kankanhalli 2008).
- **Measurement:** KM may not prove its value without measurable success (Andone 2009). Many authors have observed and proposed metrics for assess KM (Bose & Ranjit 2004; Kuah & Wong 2011; Ragab & Arisha 2013). KM implementation can be seen as an investment decision and therefore its performance outcomes must be evaluated and measured (Tabrizi et al. 2011). Measuring KM is essential in turn to ensure that its envisioned objectives are being attained or if they need to be aligned (Massingham & Massingham 2014), and tracks the progress of KM and to determine its benefits and effectiveness (Chong & Chong 2009; Migdadi 2009). Suitable KM performance measurement tool should be able to indicate opportunities for improve KM programs (Kuah & Wong 2011).

- **Implementation:** This activity implies to carry out designed KM projects and follow KM process pattern. For success of implementing KM project, it has to be in line with the political aspects, and also consider key organizational elements (Syed-Ikhsan & Rowland 2004).

**RESEARCH QUESTIONS AND INCLUSION AND EXCLUSION CRITERIA**

In this section, Research Questions (RQs) were formulated following the recommendations proposed by Kitchenham (2007). Figure 2, illustrates Research Questions (RQ1: Benefits and RQ2: Quality Attributes) and the criteria to include (I) or exclude (E) studies.

| RQ1: What are the benefits (B1 to B5) of the proposals to knowledge measurement in the public organizations? | B1: Alignment of knowledge management with strategic objectives |
| --- | --- |
| | B2: Reduction of corruption rate |
| | B3: Reduction of labor lawsuits |
| | B4: Reduction of time and cost |
| | B5: Improvement of citizen perception |
| RQ2: There have been some contributions to understandability, clarity and credibility for public organizations? | QA1: Understandability |
| | QA2: Clarity |
| | QA3: Credibility |

**Inclusion (I) and Exclusion (E) criteria**

- **I1**: The study was conducted in a public organization or can be applied in the same
- **I2**: The study contributes to the improvement of some organizational process
- **E1**: If two papers publish the same study, the less mature one is excluded
- **E2**: If the study presented no evidence, is excluded

Figure 2: Research questions and inclusion and exclusion criteria

**SEARCH PROCESS**

We design a SLR protocol to guide the search process. Relevant papers are retrieved automatically from the electronic resources, as well as from target journals, conferences, and workshops as a supplementary source to search. We set search terms based on the following keywords: Knowledge Management, Measurement, Public Administration, Public Sector, e-Government and Strategic Objectives. We agreed to accept articles in three languages (English, Spanish and Portuguese). The period of SLR covers from 2005 until 2014. We performed our database search using logical command "AND" for chain of two keywords looking to have as much as possible relevant information for answer our RQs. The study selection process was performed in three phases.

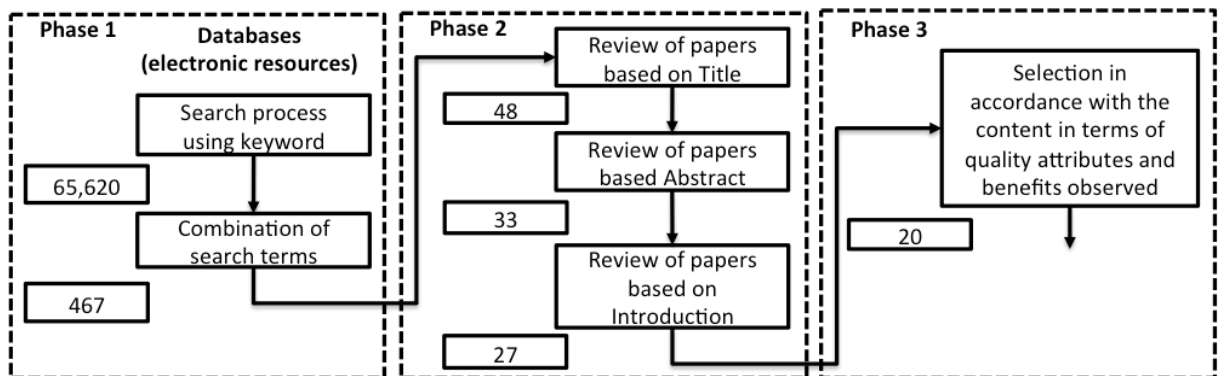

Figure 3: Search process

In short, as can be seen in figure 3, in phase 1 we conduct research through databases and we found up to 65,620 papers. Then, refining search terms we obtained 467 papers. Consequently, in phase 2, we filter by title (48), then by abstract (33), and finally based on introduction to obtain 27 potentially relevant studies. Finally, in phase 3, according with benefits and quality attributes expected, we selected the last 20 papers for SLR.

**RESULTS OF SLR**

The following paragraphs summarizes the answers found:

**RQ1**: What are the benefits (B1 to B5) of the proposals to knowledge measurement in the public organizations?

- **B1: Alignment of knowledge management with strategic objectives**. KM implementation provides information to apply managerial control systems that let organization to regard at the fulfillment of goals (Dalkir et al. 2007), thought measurement of strategic plans, monitoring achievement of its strategic objectives (Massingham & Massingham 2014).
- **B2: Reduction of corruption rate.** Using monitoring tools KM can reduce corruption by means of transparency and accountability (Tung & Rieck 2005).
- **B3: Reduction of labor lawsuits.** KM can reduce time and cost of labor lawsuits, by refining internal processes and exploitation of knowledge sources, affecting efficiency and effectiveness in POs (Fersini et al. 2013).
- **B4: Reduction of time and cost**. KM benefits time and cost reduction by implementing KM processes (Lee et al. 2005) and technological tools (Liberona & Ruiz 2013); and developing capable workers (Mbhalati 2014). KM potentially improves productivity (Choy et al. 2006).
- **B5: Improvement of citizen perception**. KM improves citizen perception as a result of improving efficiency (Savvas & Bassiliades 2009) and quality of public services (Anantatmula & Kanungo 2007; Brito et al. 2012; Mbhalati 2014).

**RQ2**: There have been some contributions to understandability, clarity and credibility for public organizations?

- *QA1: Understandability.* KM process improves KM understandability in organization thus a comprehension of KM core activities and its effects on organizational performance. Process and its activities help people and organization to better understand the purpose of KM initiatives (Lee et al. 2005; Choy et al. 2006; Ho et al. 2014) and its strategically implementation. This quality attribute is related with clarity.
- *QA2: Clarity.* KM implementation provides a better clarity about roles, tasks, and responsibilities (Puron-Cid 2014). Moreover, clarity of purpose may be enhanced due to provide standardized and documented KM policies and procedures to ensure clarity of roles and processes (Ragab & Arisha 2013). Also, through measurement organization gains clarity about what organization needs to redefine on its KM process (Bose & Ranjit 2004).
- *QA3: Credibility.* KM and its evaluation offer greater credibility about results for organizational stakeholders, and helps organization to regard at goals achievement (Dalkir et al. 2007). Credibility may be enhanced due to better quality of produced knowledge (Torres-Narváez M et al. 2014), and gives more credibility to the knowledge that they produce (Dalkir et al. 2007).

**STUDY QUALITY ASSESSMENT**

For assessing the quality of this SLR, we predominantly focus on the level of evidence as a criteria for study quality assessment used by Ali et al. (2010) and Dybå & Dingsøyr (2008). The quality assessment on the selected studies is also useful to increase the accuracy of the data extraction result. In general, we assume that overall quality of the selected studies is acceptable. We finish an acceptable SLR with a total quality factor of 4.43 with a scale of 1 to 5, as presented on figure 4.

| Study | Reference | Q1 | Q2 | Q3 | Q4 | Q5 | Total Score |
|---|---|---|---|---|---|---|---|
| S1 | (Pardo et al. 2013) | 1 | 1 | 1 | 1 | 0.5 | 4.5 |
| S2 | (Dalkir et al. 2007) | 0.8 | 1 | 1 | 1 | 1 | 4.8 |
| S3 | (Braun & Mueller 2014) | 1 | 1 | 1 | 1 | 0.5 | 4.5 |
| S4 | (Fersini et al. 2013) | 0.8 | 1 | 1 | 1 | 0.5 | 4.3 |
| S5 | (Liberona & Ruiz 2013) | 0.6 | 1 | 1 | 1 | 0.5 | 4.1 |
| S6 | (Brito et al. 2012) | 0.8 | 1 | 0.5 | 0.5 | 0 | 2.8 |
| S7 | (Torres-Narváez M et al. 2014) | 0.8 | 1 | 0.5 | 1 | 0.5 | 3.8 |
| S8 | (Migdadi 2009) | 1 | 1 | 1 | 1 | 1 | 5 |
| S9 | (Alsadhan et al. 2008) | 1 | 1 | 1 | 1 | 1 | 5 |
| S10 | (Mas-Machuca & Martínez Costa 2012) | 0.8 | 1 | 1 | 1 | 1 | 4.8 |
| S11 | (Tabrizi et al. 2011) | 0.8 | 1 | 1 | 1 | 1 | 4.8 |
| S12 | (Pee & Kankanhalli 2008) | 0.8 | 1 | 0.5 | 1 | 0 | 3.3 |
| S13 | (Anantatmula & Kanungo 2007) | 0.8 | 1 | 1 | 1 | 1 | 4.8 |
| S14 | (Akhavan et al. 2006) | 0.8 | 1 | 1 | 1 | 1 | 4.8 |
| S15 | (Choy et al. 2006) | 0.8 | 1 | 1 | 1 | 1 | 4.8 |
| S16 | (Ragab & Arisha 2013) | 0.6 | 1 | 0.5 | 1 | 0 | 3.1 |
| S17 | (Ho et al. 2014) | 1 | 1 | 1 | 1 | 1 | 5 |
| S18 | (Puron-Cid, 2014) | 1 | 1 | 1 | 1 | 1 | 5 |
| S19 | (Lee et al. 2005) | 1 | 1 | 1 | 1 | 0.5 | 4.5 |
| S20 | (Massingham 2014) | 0.8 | 1 | 1 | 1 | 1 | 4.8 |
| | **Average Score** | 0.85 | 1.00 | 0.90 | 0.98 | 0.70 | |
| | **Total Score** | 4.43 | | | | | |

Q1: Level of evidence
- L0: 0 - No evidence
- L1: 0.2 - Evidence obtained from mathematical demonstrations
- L2: 0.4 - Evidence obtained from expert opinions or observations
- L3: 0.6 - Evidence obtained from academic experiments
- L4: 0.8 - Evidence obtained from industrial studies
- L5: 1.0 - Evidence obtained from industrial practices

Q2: Is it clear the benefit of this study?
Q3: Are indicators clearly used in the study?
Q4: Are clear improvements to Quality Attributes?
Q5: Are explicitly discussed the limitations in the study?

Q2..Q5: 1=Yes; 0.5=Partially; 0=no

Figure 4: Quality assessment of the SLR

## 3. STRATEGIES TO ENABLE KM PROGRAMS IN POs

In order to identify how knowledge is managed in Mexican POs, we obtained from an ongoing project the evaluation results of the KM factors ranked in cultural, technological and strategic dimensions. Among the 32 States of Mexico, the state of Guanajuato was selected for this project mainly because its economic growth in 2014 (IMCO 2014; Milenio 2014). In this project, experts use the KM initial diagnosis (KM-id) tool defined in the Promise Framework (Mora Soto 2011) to evaluate the KM factors and get the cultural, technological and strategic difficulties before to plan and implement a KM program. Results came from five different POs in Guanajuato, Mexico.

On table 1, we present some strategies for fostering KM programs. Strategies have been aligned with dimension and difficulties or issues and its possible benefits. Issues for Mexican POs are market in italics.

**Table 1.** Strategies for fostering KM programs

| Dimension | Difficulty or issue | Strategies | Possible benefits |
|---|---|---|---|
| **Culture** | 1. Weak trust, collaboration, confidence or personnel support<br>2. Unsatisfactory capabilities or personnel profile backgrounds or trainings<br>3. *Duplicity of roles and activities*<br>4. *Limited knowledge reuse* | (1, 4) Communities of practice, team buildings and foster leadership in KM programs (S2), foster knowledge dissemination among organization for avoiding knowledge concentration in a few positions (S6) and provide pseudonyms for more privacy in KM systems.<br><br>(3) Promote norms and policies for KM (S16) cooperation and reciprocity (S12), Workshops and conferences for knowledge sharing and create an environment for KM (S10).<br><br>(2) Rewards, recognitions or benefits for personnel activities related with knowledge sharing (tied to KM strategic objectives) (S6, S16).<br><br>(2) Capabilities mapping (S20) and Improve personnel capabilities (S2, S3, S18) and develop ad-hoc training programs (S14) and technology transfer<br>For additional contribution to Mexico, leaders need to incentive trust, collaboration and self confidence among staff. | Knowledge sharing, Knowledge quality improvements<br>Enhance knowledge usage and reuse<br>Enrichment of knowledge quality attributes<br>Avoid bureaucratic discretion decision-making<br>Improve citizen perception and credibility of information resources. |
| **Technology** | 1. Not proper KM systems, difficult of use, or not interconnected nor inter-operable systems.<br>2. Insufficient ICT infrastructure<br>3. Repetitive or nonsense processes, and unlinked and untraceable activities<br>4. *Lack of standardized processes and communication channels*<br>5. *Lack of measurement and control of organizational knowledge assets*<br>6. *Absence of homogenized technological support implementation in POs*<br>7. *Many core activities must be manually performed; consequently processes execution become slow, time consuming, and expensive.* | (1, 6) Ease of use and access technologies and systems design (S1), based on users experience (S1), KM systems evaluation trough Knowledge Quality Attributes (KQA) (S1, S18), integrate informational retrieval systems (S4), knowledge ontology for knowledge representations (S4), KM systems inclusion of meta-data, taxonomy, thesauri, document/contact management, learning support systems, and knowledge repositories (S12), variety of generic KM tools for daily activities (S12, S16), guarantee connectivity (S20) and KM activities systematic integration (S15).<br><br>(4) Define standards and protocols for Inter-operability among POs (S1, S2) and build relationships with other institutions (S2).<br><br>(3, 4, 7) Identify key knowledge and core process (S5, S17, S19) and process (S18, S19) and documents (S4, S16) standardization, and organizational process re-engineering (S14, S18).<br><br>(5) Measurement of KM activities effects on POs (S10). Nurture strategic | Facilitate KM system operation<br>Improve knowledge sharing,<br>Better decision making process,<br>Improve public policies<br>Improve efficiency and effectiveness<br>Cost reductions and effort diminutions<br>Inter-operation of KM Systems. |

|  |  | measurement for KM activities (S6), continuum improvement of process quality based on performance standards (S7, S19).<br><br>(2) Platforms as a service with software assurance. |  |
| --- | --- | --- | --- |
| **Strategy** | 1. Weak vision, confused purpose<br>2. Political changes and not inter-administration continuity<br>3. Lack of political willingness<br>4. Lack of leaderships and KM support<br>5. *Organizational statutes do not encourage value creation, use and reuse of organizational knowledge assets*<br>6. *Misalignment of effective knowledge measurement.* | (2, 3, 4, 5, 6) Top management support (S3, S4, S8, S9, S14, S18) and commitment (S5), legislative support (S18), ICT policies and standards (S18), and improve POs awareness about KM benefits (S5), and guarantee financial resources for KM implementation (S5).<br><br>(1, 3) Establish KM strategy (S9, S12, S14, S18) and organizational change program (S20), strategic focus, measurement of results, content quality and collaboration (S13), link KM strategy to institution global strategy (S16) and align KM with organizational resources and budget (S9), and communicate its benefits (S16) among the organization. Adopt a long term KM vision (S9, S12).<br><br>(1) Assign personnel for KM initiatives (S5) and better staff training and development (S11).<br><br>(6) Supervision of KM enablers (monitoring) (S17) and measure KM implementation-program (S17) and internal outcomes and benefits (S17), consider future requirements (people, systems, process) (S20).<br><br>(1) Promote a shared vision about KM intention.<br><br>(6) Establish political objectives alignment with KM initiatives.<br><br>(1, 2, 3) Report KM benefits in ways that politicians understand, recognize and appreciate short and long term benefits.<br><br>(2, 3, 4) Explain purpose, scope, roles, and cognition on KM to POs stakeholders. | Improve KM implementation outcomes<br>Help to improve KM processes<br>Foster innovation and KM quality improvements<br>Guarantee KM initiatives continuity. |

## 4. DISCUSSIONS

Finally, we consider that knowledge reuse is highly related with the **cultural** dimension, associated with trust, transparency, honesty, collaboration, professionalism, flexibility, commitment, learning, innovation and expertise. The use of existing knowledge base is an elemental practice for a good administration and knowledge reuse (Savvas & Bassiliades 2009). As empirical research have shown, KM can minimize duplication of efforts and promote services consistency (Lee 2008). An organizational culture that supports sharing, using and reusing of knowledge has been identified as a enabler for KM (Choy 2006). Therefore, an environment where people do not feel forced to share knowledge, but where they have a constant desire to learn together for complementing each other is needed.

**Technological** dimension involves measurement, business process and technological infrastructure. Technological infrastructure provides a control base for assessing KM initiatives success in order to detect opportunities for realignment organizational KM strategy (Akhavan et al. 2006). KM measurement activities give a sort of strategic control for managers to improve decision-making and for regulate knowledge asset usage, is important to have a KM evaluation plan that assess KM value (Migdadi 2009). Through this it is possible to detect what kind of technical skills and infrastructural needs are necessary for KM initiatives.

**In strategic dimension,** public policy, seen as legislation and normative frameworks, represent a knowledge enabler for developing appropriate and wide KM initiatives that encourage KM activities like sharing and program evaluation. Executive level leadership represented as support of policy makers is needed in order to develop ICT-related policies. This should be legitimated by the connection between knowledge management and organizational objectives and a strong supplement of political goodwill (Mbhalati 2014) for this initiatives. Problems of missing legal, statute or normative framework supporting KM initiatives, can be caused by ignorance about the relevance and benefits of KM (Gil-García & Pardo 2005). For these reasons institutional and organizational factors like norms, regulations, resources, plans, standards, processes, level of centralization, and jurisdictions are so relevant for a better adoption of KM initiatives.

**Acknowledgments**

This research was partially funded by the National Council of Science and Technology of Mexico (CONACyT) through the project "Labor Justice System Development for the State of Guanajuato" (FOMIX-219762).